\documentclass[preprint]{elsarticle}
\usepackage{tabularx} 
\usepackage{amsmath}  
\usepackage{amsfonts} 
\usepackage{graphicx} 
\usepackage{xcolor}
\usepackage[margin=1in,letterpaper]{geometry} 
\usepackage[final]{hyperref} 
\hypersetup{
    colorlinks=true,       
    linkcolor=blue,        
    citecolor=blue,        
    filecolor=magenta,     
    urlcolor=blue   
}

\newcommand{\m}{\mu}
\newcommand{\n}{\nu}
\def\r{\rho}
\def\s{\sigma}
\def\l{\lambda}

\def\a{\alpha}

\def\d{\delta}
\def\k{\kappa}
\def\c{\chi}

\begin{document}

\title{A Generally Covariant Model of Spacetime as a 4-Brane in 4+1 Flat Dimensions}
\author[1]{Mert Ergen}
\ead{mert.ergen@boun.edu.tr}
\author[2]{Metin Arık}
\ead{metin.arik@boun.edu.tr}
\affiliation[1]{organization={Bogazici University}, addressline={Bebek}, postcode={34342}, city={Istanbul}, country={Turkiye}}
\affiliation[2]{organization={Bogazici University}, addressline={Bebek}, postcode={34342}, city={Istanbul}, country={Turkiye}}
\date{}
\begin{abstract}
We construct a simple toy model of a closed FLRW spacetime by starting from a flat five dimensional scalar field space. The action contains no curvature terms. A string-like potential is used for the field space. The model is fully covariant and the spacetime metric is uniquely determined from the dynamical equations of the metric tensor. An SO(4) invariant ansatz for the scalar fields is shown to be sufficient to build the FLRW cosmology.
\end{abstract}
\maketitle

\section{Introduction}
General covariance is probably the most appealing aspect of Einstein's theory of general relativity. General relativity was built on the foundation of special relativity. Since spacetime in special relativity is flat, incorporating gravity into the theory was accomplished through the curvature concept and the Riemann tensor. This historical order of events can make one think that the action of the theory must contain some form of the Riemann tensor, tied to the dynamics of the metric. However, it is possible to construct a fully covariant theory without using any derivatives of the metric in the action. We do not necessarily have to work with curvature terms. We know from the works of Sakharov concerning the quantum fluctuations on the curvature\cite{sakharov67,visser02} that different orders of curvature may be derived as quantum corrections to the action. Non-minimal couplings with the curvature term are also viable\cite{sushkov09}. There are also works that connect scalar field extensions to spacetime with modified gravity theories and produce similar cosmological results\cite{briscese07,nojiri07,nojiri07_2,capozziello06}. Scalar fields are also used to describe inflation, as has been investigated in various papers\cite{wang20,piran85,liddle98,belinsky88}.

\par
Extending the configuration space through physical or internal degrees of freedom has been thoroughly used since Kaluza-Klein as a general technique. Cosmological studies also utilize this method\cite{arik89,sengor13,tiwari10}. Several papers\cite{ratra_inflation92,gottlöber_inflation91,elizalde_inflation08,capozziello_inflation06} use scalar fields in flat or curved spaces to define inflation fields. Palatini approach to general relativity also utilizes scalar fields\cite{bauer11,kaewkhao18}. Our method will also have similarities with the Polyakov action and investigations stemming from Polyakov's work.\cite{polyakov81}
\par 
This work aims to build on the aforementioned concepts to obtain a curved spacetime obeying FLRW geometry as the ground state of the expanding universe. The use of scalar fields is taken in likelihood of the Higgs mechanism of the Standard Model. This toy model is aimed to explore a new idea to see if established knowledge coming from the Einstein theory can be reached through other means.
\par
We make the usual assumptions of homogeneity and isotropy for the cosmological universe. As explained, we also want the universe to expand, in accordance with modern astronomical observations so we choose the most appropriate metric as the FLRW metric. The metric that we want as the ground state solution of our theory is

\begin{align}\label{eq1}
\mathrm{g} = dt^2 - a^2 \mathrm{g}_\Sigma.
\end{align}

where the speed of light $c$ is set to unity, $a$ is the scale size and $\mathrm{g}_\Sigma$ is a 3-dimensional space with uniform curvature. We prefer the homogeneous and isotropic universe starting from a point. The term $\mathrm{g}_\Sigma$ contains

\begin{align}
\mathrm{g}_\Sigma = \frac{dr^2}{1 - kr^2} + r^2 \mathrm{g}_{S^2}
\end{align}

where $r$ can be thought of as the radius of spacelike sections and $\mathrm{g}_{S^2}$ is the metric of $S^2$. In this convention, $k$ has the dimensions of $length^{-2}$ and is the Gaussian curvature of this space. It determines if the spacelike sections of the space will be closed, flat or open. In this paper we prefer to work with the convention that keeps $r$ and $k$ dimensionless so the scale size will have the $length$ dimension. An important remark that will be expanded later is that we want the $k = 1$ closed universe with positive constant curvature choice for the spacelike sections of the metric.
\par
The scale size $a(t)$ multiplies the spatial part of the metric and determines the expansion of this space in time. The present time in this setup is $a(t_0) = 1$ while the theoretical Big Bang resides at $a(0) = 0$. An observation here is that the scale size does not change the solid angle but governs the radial expansion. We will manipulate this form of the line element in the next section. 

\section{Cartesian Coordinates for FLRW Metric with $k = 1$}

The theory of general relativity and the cosmological observations lead us to make the homogeneity and isotropy assumptions. The homogeneity is an observational fact that can be seen through cosmic background radiation and similar evidences, which enforces the idea that there are no ``special'' places in the universe contrary to the geocentric models. The isotropy claim is similarly observationally apparent: the cosmic microwave background temperature fluctuations are small throughout the universe. Also as we, at an arbitrary point in the universe, observe no special direction it seems plausible that no such direction exists for any point. This line of thought also leads to homogeneity\cite{straumann74}.
\par
The simplest line element with these assumptions is the FLRW metric for a closed universe as stated before. This line element can be sliced in the time dimension to acquire an $SO(4)$ invariant space. We start with the generic FLRW metric

\begin{align} \label{FLRW generic}
\mathrm{g} = dt^2 - a^2(t) \frac{dx^2 + dy^2 + dz^2}{(1 + \frac{k}{4}(x^2 + y^2 + z^2))^2}
\end{align}

where we take $t$ as the cosmological time, $a(t)$ as the scale size that depends only on the cosmological time and governs the expansion of the universe. $k = 1$ for the closed and constant-curvature spacelike sections. In this setup we look at the line element:

\begin{align}\label{line_element}
\mathrm{g}_ \Sigma = \frac{dx^2 + dy^2 + dz^2}{(1 + \frac{1}{4}r^2)^2} = d\phi^2 + sin^2(\phi) \mathrm{g}_{S^2}.
\end{align}

The rightmost side of the equation brings to mind the generic spherical setup. Thus for $k = 1$ FLRW metric we can write

\begin{align}
\mathrm{g} = \left(\frac{dt}{da}\right)^2 da^2 - a^2\mathrm{g}_ \Sigma 
\end{align}

where we think of the cosmological time as dependent on the scale size and the scale size itself as independent. We assume $a(t)$ is a smooth function of time, excluding possible singularities. Rewriting,

\begin{align}
\mathrm{g} = \frac{1}{\dot{a}^2} da^2 + da^2 - (da^2 + a^2\mathrm{g}_ \Sigma).
\end{align}

We add and subtract a term to shape the negative part of the line element into the metric of $\mathbb{R}^4$ in hyperspherical coordinates. We define the Cartesian coordinates as

\begin{align}
&x^0 = a cos\c \\
&x^1 = a sin\c sin\theta cos\phi \\
&x^2 = a sin\c sin\theta sin\phi \\
&x^3 = a sin\c cos\theta.
\end{align}

Note that these coordinates $x^\m$ can be related to the coordinates in (\ref{FLRW generic}) but are different from them. Also note that although we are using $x^0, x^1, ...$ as coordinates, the metric for the coordinates is not Minkowskian $\eta_{\m\n}$ but Euclidean $\d_{\m\n}$. Using these coordinates we treat the spatial part of the metric as the flat $\mathbb{R}^4$, so that

\begin{align}
\mathrm{g} = \left( 1 + \left(\frac{dt}{da}\right)^2 \right) da^2 - \d_{\m\n} dx^\m dx^\n.
\end{align}

where the last term is the flat 4-dimensional metric. Looking at the $a^2$ term, we can express the line element:

\begin{align}
&a^2 = x^2 = \d_{\m\n} x^\m x^\n \\
&a da = \d_{\m\n} x^\m dx^\n = x_\n dx^\n \\
&\mathrm{g} = \left( 1 + \left(\frac{dt}{da}\right)^2 \right) \frac{x_\m x_\n dx^\m dx^\n}{a^2} - \d_{\m\n} dx^\m dx^\n.
\end{align}

Now we can isolate the metric tensor from here

\begin{align}
&\mathrm{g} = g_{\m\n} dx^\m dx^\n \\
&g_{\m\n} = (1 + \l) \frac{x_\m x_\n}{x^2} - \d_{\m\n}.
\end{align}

We now find the eigenvalues of the metric.

\begin{align}\label{evalue}
g_{\m\n} x_\n = \l x_\m && g_{\m\n} y_\n = -y_\m.
\end{align}

Here the summation convention over repeated indices is used but the equations are purely algebraic. $x_{\m} = x^{\m}$ is not a covariant vector in the curved Riemannian geometry but denotes the algebraic eigenvector of the matrix $g_{\m\n}$ corresponding to the eigenvalue $\l$. $y_\n$ is similar in nature. We want $\l$ to be positive to correspond to the Minkowskian signature and we want $y_\n$ to be  orthogonal to $x_\n$ in a Cartesian sense so we have eigenvalue -1 using $x_\n y^\n = 0$.
\par
This manipulation is an important aspect of our calculation. The generators of $\mathrm{SO(4)}$ cannot act on the coordinates of $S^3$ in a linear way but they act linearly on these Euclidean coordinates. The importance of the FLRW $k = 1$ metric is that the symmetry group $\mathrm{SO(4)}$, under which FLRW metric is invariant, acts linearly on these coordinates. For FLRW $k = 0$ metric, on which the Euclidean group acts as the invariance group, the covariant coordinates are the $(x, y, z)$ tuple in (\ref{FLRW generic}). For $k = -1$, the desired covariant coordinates on which $\mathrm{SO(3,1)}$ acts linearly are given by equations similar to (\ref{evalue}) with the Euclidean metric $\d_{\m\n}$ replaced with the Minkowskian metric $\eta_{\m\n}$.

\section{Proposed Lagrangian Density}

We start with flat 4+1-dimensional configuration space which has the usual Minkowskian signature, so $f_{ab} = diag(+1, -1, -1, -1, -1)$. Our aim is to place the simplest field in this space and use its dynamics to generate the metric for 3+1-dimensional FLRW spacetime. We will call this field $\phi(x)$ and expect it to be invariant under $SO(4)$ acting on the $x$-coordinates, satisfying the FLRW expanding universe conditions. This approach assumes that the universe is closed and and the spatial sections have spherical symmetry. 
\par
Our method places the emphasis on the fields. We believe this has ontological significance in the sense that the dynamics of the underlying field defines the structure of the spacetime instead of an ad hoc metric.

The Lagrangian density in this case is

\begin{align}
\mathcal{L} = \sqrt{|g|} \left[ \frac{1}{2} g^{\m\n} f_{ab} \partial_\m \phi^a \partial_\n \phi^b - V(\phi) \right]
\end{align}

where $a,b = 0, 1, 2, 3, 4$; $\m,\n = 0, 1, 2, 3$; $g^{\m\n}$ is the inverse metric of spacetime, $f_{ab}$ is the metric of the configuration space and the potential $V(\phi)$ is left to be determined. We choose $f_{ab} = diag(+, -, -, -, -)$ as the metric of the 5-dimensional configuration space. Looking at the Euler-Lagrange equations for the metric inverse 

\begin{align}\label{E-L}
& \frac{\partial \mathcal{L}}{\partial g^{\rho\sigma}} = - \sqrt{|g|} \left[
 \left(
  \frac{1}{2} g_{\r\s} g^{\m\n} f_{ab} \partial_\m \phi^a \partial_\n \phi^b - V(\phi) \right) + \left(\frac{1}{2} f_{ab} \partial_\rho \phi^a \partial_\sigma \phi^b\right) \right] = 0
\end{align}
since
\begin{align}
& \frac{\partial \mathcal{L}}{\partial (\partial_\mu g^{\rho\sigma})} = 0.
\end{align}

(\ref{E-L}) can be solved for the spacetime metric.

\begin{align}
&g_{\m\n} ( g^{\r\s}f_{ab} \partial_\r\phi^a\partial_\s\phi^b - V) = f_{ab} \partial_\m\phi^a\partial_\n\phi^b \\
&\Phi_{\m\n} := f_{ab}\partial_\m\phi^a\partial_\n\phi^b, \quad \Phi := g^{\m\n}\Phi_{\m\n} \\
&g_{\m\n}(\Phi - V) = \Phi_{\m\n} \label{phiV} \\
&g_{\m\n} = \frac{\Phi_{\m\n}}{\Phi - V}
\end{align}
and we can contract (\ref{phiV}) with the inverse metric to get

\begin{align}
4(\Phi - V) = \Phi
\end{align}

so we can conclude

\begin{align}
g_{\m\n} = C.\frac{\Phi_{\m\n}}{V}
\end{align}
where $C$ is a dimensionless constant. We thus can find the metric explicitly in terms of the scalar fields. Also note that these equations point to a zero total energy-momentum tensor for the entire universe. This is a significant assumption related to the foundation of our theory and has been explored before\cite{rosen94, johri95}. 

For sake of clarity, we now change our notation to distincly show the fields and the solution for the metric is

\begin{align}
g_{\m\n} = \frac{\partial_\m \c \partial_\n \c - \partial_\m \phi^\a \partial_\n \phi^\a}{V}
\end{align}

where $\a = 1 , 2, 3, 4$ and $\phi^0=\c$. Now that we have the expression for the metric, we want to solve the Euler-Lagrange equations of the scalar field 

\begin{align}\label{Lphia}
\frac{1}{\sqrt{|g|}} \partial_\m \left(\sqrt{|g|} g^{\m\n} \partial_\n \phi^\a\right)
= - \frac{\partial V}{\partial \phi^\a} \\
\frac{1}{\sqrt{|g|}} \partial_\m \left(\sqrt{|g|} g^{\m\n} \partial_\n \c\right)
= - \frac{\partial V}{\partial \c}
\end{align}
where $\a = 1, 2, 3, 4$ becomes a Euclidean index. 

\subsection{Scale Invariant Assumption For The Action}

We define the potential

\begin{align}
&V = \frac{1}{4}\l\phi^4 + \frac{1}{4}\k \c^4 + \frac{1}{2} \n \phi^2\c^2.
\end{align}

This is a standard interaction potential that will keep the action scale invariant together with our assumptions for the fields. We set our assumption for the $\phi$-field as

\begin{align}
\phi^\m = C \frac{x^\m}{x^2}
\end{align}

where $C$ is a dimensionless free constant. This also ensures the scale invariancy of the field. When we choose the potential and the  $\phi$-field as such, the scalar field must have this corresponding form

\begin{align}
\c = \frac{D}{\sqrt{x^2}}
\end{align}

where $D$ is also a dimensionless constant. Now that all of the fields in the action are scale invariant, we can move along with the calculations. With these assumptions, the metric takes the form

\begin{align}
g_{\m\n} = \frac{3C^2}{Vx^4}(- \d_{\m\n} + \frac{C^2 + D^2}{C^2}\frac{x_\m x_\n}{x^2})
\end{align}

where $V$ is the potential defined above. We can see that the metric is like $\d_{\m\n}$ plus corrections and it is locally Minkowskian. Using the metric identity we can get the metric inverse

\begin{align}
g^{\m\n} = \frac{Vx^4}{3C^2} (- \d^{\m\n} + \frac{C^2 + D^2}{D^2} \frac{x^\m x^\n}{x^2})
\end{align}

and with the assumptions put in, the metric determinant is

\begin{align}
\sqrt{|g|} = \frac{3CD}{Vx^4}.
\end{align}

Rewriting the potential in terms of the fields we have proposed, we have

\begin{align}
V = \frac{1}{4x^4}\l C^4 + \frac{1}{4x^4}\k D^4 + \frac{1}{2x^4} \n C^2D^2 = \frac{\m}{4x^4}
\end{align}

where $\k, \l, \n, \m$ are all dimensionless. The parameter $\m := \l C^4 + \k D^4 + 2\n C^2D^2$ is defined for brevity.
The next step would be to get the resulting Euler-Lagrange equations:

\begin{align}
\frac{1}{6} (\l C^4 + \k D^4 + 2\n C^2 D^2) &= \l C^4 + \n C^2 D^2 \\
\frac{1}{12} (\l C^4 + \k D^4 + 2\n C^2 D^2) &= - (\k C^2D^2 + \n C^4).
\end{align}

Solving these yield the necessary relations as

\begin{align}
\k = \frac{C^4}{D^4} \l \quad , \quad \n = -\frac{C^2}{D^2} \l.
\end{align}

Using these relations we can see that a potential of this form will vanish:

\begin{align}
V = \frac{1}{4x^4} C^4 \l + \frac{1}{4x^4} D^4 (\frac{C^4}{D^4} \l) 
 - \frac{1}{2x^4}  C^2D^2 (\frac{C^2}{D^2} \l) = 0
\end{align}

Our metric has the potential expression in the denominator. Since it appears in the denominator we have to change it to prevent escaping to infinity. We redefine the potential to have a new form

\begin{align}
V = \frac{1}{4} \l \phi^4 + \frac{1}{4} \k \c^4 + \frac{1}{2} \n \phi^2\c^2 + \frac{1}{2}m_\c^2\c^2
\end{align}

where $m_\c$ has the dimensions of $mass$. Using no mass term was our first aim as to preserve the $\mathnormal{SO(4)}$ symmetry and also keep the model as simple as possible with our field and potential forms but this occurance of identically zero potential nudges us to break the symmetry have thus far preserved. Still abstaining from complicating the model, we first add a mass term to only one of the fields we have. Solving the new Euler-Lagrange equations we find
that $m_\c^2$ should be equal to 0 for this system to be solved. Adding a mass term only to the remaining four scalar fields does not work either, with the similar solution $m_\phi^2 = 0$.
The model forces us to break the symmetry with two different masses related to the fields.
When we add both terms to the potential it becomes

\begin{align}
V = \frac{1}{4} \l \phi^4 + \frac{1}{4} \k \c^4 + \frac{1}{2} \n \phi^2\c^2 + \frac{1}{2}m_\phi^2\phi^2 + \frac{1}{2}m_\c^2\c^2.
\end{align}

This can also be manipulated to show a certain aspect of the potential:

\begin{align}
&V = (\frac{\sqrt{\l}}{2} \phi^2 - \frac{\sqrt{\k}}{2}{\c^2} - m^2)^2 - m^4
\end{align}

where $m$ has the dimensions of mass and used as a constant term representing the mass scale. The form of our potential should bring to mind the Higgs mechanism. Although we have drawn the inspiration from there, what we are doing here is not minimizing a potential at all times and perturbing the field around a minimum to get a Higgs particle. In fact, our construction is completely based on cosmological assumptions so it is not quantum at all. Our solution is in curved space from the start. To find the particles related to these fields and turn this theory into a quantum theory, we can safely say that we are working in flat Minkowskian space since we are interested in short-range interactions only. Then the usual procedure for quantum field theories and renormalization can be followed. Contrary to the Higgs mechanism, our perturbation around a minimum would be dependent on cosmological time and so subject to change. So it is apparent that this theory only resembles Higgs theory on the surface.

The Euler-Lagrange equations calculated with this potential give meaningful results:

\begin{align}
&\phi^\m: \frac{\m + 2m_\phi^2C^2X^2 + 2m_\c^2D^2X^2}{12}\left(\frac{C^2}{D^2} - 3\right) = - \left(\l C^4 + \n C^2D^2 + m_\phi^2C^2X^2\right) \\
&\c: \frac{\m + 2m_\phi^2C^2X^2 + 2m_\c^2D^2X^2}{12} = \k D^4 + \n C^2D^2 + m_\c^2D^2X^2.
\end{align}

We can now find a relation between $m_\phi$ and $m_\c$

\begin{align}\label{mass_relation}
\frac{m_\c^2}{m_\phi^2} = \frac{C^2}{D^2} \frac{15D^2 - C^2}{C^2 - 3D^2}
\end{align}

Putting this relation and the field assumptions into the potential we acquire

\begin{align}
V = \frac{1}{4X^4} \left(\m + m_\phi^2C^2X^2 \frac{C^2 + 9D^2}{C^2 - 3D^2} \right)
\end{align}

where $\m := \frac{1}{4}(\l C^4 + \k D^4 + 2\n C^2D^2)$ which is still identically zero since the massive part of the equations does not constrain the relations of the massles parameters. We thus have a minimum for the potential and a plausible theory.
\par

When we plug in the new potential and the mass relation to the closed form of the metric we get

\begin{align}
g_{\m\n} = \frac{\frac{C^2}{D^2} - 3}{2m_\phi^2X^2} \left( -\d_{\m\n} + (1 + \frac{D^2}{C^2}) \frac{X_\m X_\n}{X^2} \right).
\end{align}

An important observation of our theory is that the energy-momentum tensor in this universe is manifestly zero,
\begin{align}
T_{\m\n} = 0.
\end{align}

This is a result of the positive energy of the mass field balancing the negative energy of the ghost fields exactly. Some of the previous work examining this subject\cite{guth1991, bondi1997, berman2009} explain this phenomenon with the rapid expansion of the universe being powered by the transfer of energy from the gravitational field which has a negative sign. The positive energy from mass completely cancels this negative energy and the total is zero for all times. This is a more strict condition than just the divergence of the energy-momentum tensor being zero and we are persuaded this is a better explanation of the energy total of the universe since it is even less dependent on geometry and based on a more physical ground. It is also remarkable that certain equaivalent interpretations of Mach's principle(REFS), which we hold in high regard in this topic, say that the total energy, angular and linear momentum of the universe should be zero. Our theory comes to this conclusion organically.

\subsection{Constant magnitude $\phi$-field $\c$-field Assumption}

This second assumption and its results are included in the name of completeness. The ansatz and the resulting form of the $\phi$-field is

\begin{align}
\phi^\m = M\frac{x^\m}{\sqrt{x^2}} && \c = N.
\end{align}

As can be seen immediately, the magnitude of the scalar $\c$ field becomes a constant due to dimensional considerations and we lose the parameter to change the sign of the metric. This loss of parameter also corresponds to the singular nature of the metric,

\begin{align}
g_{\m\n} = \frac{M^2}{Vx^2} \d_{\m\n}.
\end{align}

This is a locally Euclidean metric and shows that this assumption cannot shape spacetime and bestow the properties we want to the metric.

\section{Conclusion}

We have shown that a theory built in a field space with only scalar fields can generate the usual expanding cosmology equipped with the FLRW metric. The model is stable with a construction that makes quantization and renormalization possible. This unique approach shows that we do not need to start from general relativity to imagine a universe aligned with the recent observations.
\par
We know from the quantization of the bosonic string that we can renormalize the theory if we have $(1+1)$-dimensional $x$-space and $(25+1)$-dimensional $\phi$-space with $V = 0$. In our theory, setting the $\phi$-space as $(n+1)$-dimensional  with the metric $f_{ab} = \eta_{ab}$ and the x-space as $(1+1)$-dimensional with $(-,+)$ signature and keeping the theory free ($V = 0$), we recover the Polyakov action for string theory\cite{polyakov81,deser76,brink76}. Our theory specifically embeds the $3+1$-dimensional spacetime into a $4+1$-dimensional flat scalar field space.
\par
The shortcomings of this theory can be summarized as such: one of our scalar fields has a negative sign which turns it into a ghost field, and the dimensional analysis of the theory shows that there is no room for a spin-2 particle that would represent the graviton. Our explanation for the ghost field, or the phantom field as preferred by the theoretical cosmologists, is that it is acting as the counterweight for the positive mass fields by driving the expansion of the universe and keeping the total energy of the universe zero. We believe that this phantom field can be mitigated using certain techniques not explored here\cite{caldwell2002, arkani-hamed2004, urban2009, ivanov2016}. The latter problem of gravitons is solved by realizing that this theory would not need spin-2 particles as the gravity is already generated by the extra scalar field.
\par
The next step for this theory would be to implement the Standard Model fields by adding them to the Lagrangian and calculating the dynamics of the interactions. Since this theory contains only scalar fields with $m^2\phi^2$ mass terms and $\phi^4$ self interactions and $\phi^2\chi^2$ interactions, we posit that at the quantum level with the aforementioned short-range considerations we will have a renormalizable field theory. We expect logarithmic divergences solvable with dimensional regularization, but we feel it best to leave this part to the quantum field theory experts. 

\section{Author Contribution}
Authors Mert Ergen and Metin Arık contributed equally for this body of work.

\section{Data Availability}
No datasets were generated or analysed during the current study.

\section{Funding Declaration}
The authors received no specific grant from any funding agency in the public, commercial, or not-for-profit sectors.

\end{document}